\documentclass[twocolumn,superscriptaddress,pr]{revtex4-1}
\usepackage{amsmath,amssymb}
\usepackage{graphicx}
\usepackage{color}

\usepackage{times}
\usepackage[colorlinks,bookmarks=false,citecolor=blue,linkcolor=red,
urlcolor=blue]{hyperref}
\begin{document}

\author{Vincenzo Alba}

\affiliation{Department of Physics and Arnold Sommerfeld 
Center for Theoretical Physics, Ludwig-Maximilians-Universit\"at 
M\"unchen, D-80333 M\"unchen, Germany}
\affiliation{International School for Advanced Studies (SISSA), Via
Bonomea 265, 34136, Trieste, Italy, INFN, Sezione di Trieste}

\date{\today}

\title{Eigenstate thermalization hypothesis (ETH) and integrability in 
quantum spin chains}

\begin{abstract} 

We investigate the eigenstate thermalization hypothesis (ETH) in 
integrable models, focusing on the spin-$\frac{1}{2}$ isotropic 
Heisenberg ($XXX$) chain. We provide numerical evidence that ETH 
holds for {\it typical} eigenstates ({\it weak} ETH scenario). 
Specifically, using a numerical implementation of state-of-the-art 
Bethe ansatz results, we study the finite-size scaling of the 
eigenstate-to-eigenstate fluctuations of the reduced density matrix. We 
find that fluctuations are normally distributed, and their standard 
deviation decays in the thermodynamic limit as $L^{-1/2}$, with $L$ the size 
of the chain. This is in contrast with the exponential decay that is found 
in generic non-integrable systems. Based on our results, it is natural to 
expect that this scenario holds in other integrable spin models and for 
typical local observables. Finally, we investigate the entanglement 
properties of the excited states of the $XXX$ chain. We numerically 
verify that typical mid-spectrum eigenstates exhibit extensive 
entanglement entropy (i.e., volume-law scaling). 

\end{abstract}

% \pacs{73.43.Cd, 71.10.Pm  {\tt check!}}

\maketitle

%\section{outline}
%\line(1,0){240}

%\line(1,0){240}

%%% INTRO 
\section{Introduction}
In classical physics the observation that at long times systems 
thermalize led to the birth of statistical mechanics as a very 
effective description of nature. However, the issue of how equilibration 
and thermalization arise in isolated quantum (many-body) systems is still 
highly debated~\cite{gemmer-2004,rigol-2007,popescu-2006,rigol-2008,
polkovnikov-2011,yukalov-2011,eisert-2014}. Recent years have witnessed a 
resurgence of interest in this topic~\cite{rigol-2007,kollath-2007,
manmana-2007,calabrese-2007,rigol-2008,cramer-2008,barthel-2008,
cramer-2008a,kollar-2008,iucci-2009,sotiriadis-2009,roux-2009,rigol-2009,
rigol-2009a,barmettler-2009,barmettler-2010,cramer-2010,flesch-2010,
roux-2010,fioretto-2010,biroli-2010,santos-2010,banuls-2011,calabrese-2011,
gogolin-2011,rigol-2011,caneva-2011,santos-2011,cassidy-2011,essler-2012,
cazalilla-2012,mossel-2012a,rigol-2012,mossel-2012,fagotti-2013,fagotti-2013a,
collura-2013,caux-2013,mussardo-2013,kormos-2013,bertini-2014,sotiriadis-2014,essler-2014,
fagotti-2014,fagotti-2014a,wouters-2014,pozsgay-2014,eisert-2014}, due to its   
relevance in cold atoms experiments~\cite{greiner-2002,kinoshita-2006,
hofferberth-2007,bloch-2008,trotzky-2012,gring-2012,cheneau-2012,
schneider-2012,kunhert-2013,langen-2013,meinert-2013,fukuhara-2013,
ronzheimer-2013,braun-2014}. One possible mechanism explaining thermalization 
is the so-called eigenstate thermalization hypothesis (ETH)~\cite{deutsch-1991,
srednicki-1994,srednicki-1996,srednicki-1999,gemmer-2004,goldstein-2006,
popescu-2006,goldstein-2010,goldstein-2010a,rigol-2008,rigol-2009,
ikeda-2011,ikeda-2013,steinigeweg-2013,beugeling-2013,zangara-2013,steinigeweg-2014,
sorg-2014,beugeling-2014,khemani-2014,kim-2014,bonnes-2014}. 

%%% ETH
The ETH can be stated as follows: In the thermodynamic limit the eigenstate 
expectation value (EEV) $\widehat{\mathcal O}_{\alpha\alpha}\equiv\langle
\psi_\alpha|\widehat{\mathcal O}|\psi_\alpha\rangle$ of a typical few-body 
observable $\widehat{\mathcal O}$ in an eigenstate $|\psi_\alpha\rangle$ 
of a many-body Hamiltonian ${\mathcal H}$ with eigenenergy $E_\alpha$ 
(and energy density $e_\alpha\equiv E_\alpha/L$, with $L$ being 
the number of sites) is equal to the microcanonical average $\langle
\widehat{\mathcal O}\rangle$, at the mean energy density $\langle{\mathcal 
H}/L\rangle=e_\alpha$, i.e.,  
\begin{equation*}
\widehat{\mathcal O}_{\alpha\alpha}=\langle\widehat{\mathcal O}\rangle,\quad
\mbox{with}\,\langle \widehat{\mathcal O}\rangle\equiv{\mathcal N}^{-1}
(e_\alpha,\Delta e)\sum\limits_{\gamma : |e_\gamma-e_\alpha|<\Delta e}
\widehat{\mathcal O}_{\gamma\gamma}.  
\label{can_aver}
\end{equation*}
Here $\Delta e$ is a small energy scale that can be sent to zero in the 
thermodynamic limit, and ${\mathcal N}(e_\alpha,\Delta e)$ denotes the 
number of energy levels $e_\gamma$ in the energy window $|e_\alpha
-e_\gamma|<\Delta e$. The underlying idea of ETH is that eigenstates with 
similar energy give similar EEVs. As a consequence, the standard deviation 
of the fluctuations, $\sigma(\widehat{\mathcal O})\equiv[\langle\widehat{
\mathcal O}^2\rangle-\langle\widehat{\mathcal O}\rangle^2]^{1/2}$, vanishes  
in the thermodynamic limit, which implies that $\widehat{\mathcal O}_{\alpha
\alpha}$ becomes a ``smooth'' function of $e_\alpha$.  

%%% FINITE SIZE SCALING 
Still, two main interpretations of ETH are possible~\cite{biroli-2010}. In the 
so-called {\it weak} ETH, ``rare'' (as opposed to {\it typical }~\cite{goldstein-2006,
popescu-2006,goldstein-2010,goldstein-2010a}) eigenstates $|\psi_r\rangle$, 
for which $\widehat{\mathcal O}_{rr}\ne\langle\widehat{\mathcal O}\rangle$, are 
allowed for any finite size. The behavior of the fluctuations reflects that the 
fraction of rare eigenstates vanishes in the thermodynamic limit. Interestingly, these 
rare eigenstates appear for Hamiltonians corresponding to random matrices~\cite{brandino-2012}. 
On the other hand, in the {\it strong} ETH these rare eigenstates are not present, i.e., all 
eigenstates are thermal. This difference is dramatically reflected in the 
finite-size scaling behavior of $\sigma(\widehat{\mathcal O})$, and it is 
related to the difference between integrable~\cite{caux-2011} and non integrable 
models. In the latter the strong ETH holds, and it is now well established that 
fluctuations decay exponentially with system size~\cite{rigol-2008,santos-2010,
rigol-2012,steinigeweg-2013,steinigeweg-2014,beugeling-2013,steinigeweg-2014,
sorg-2014,kim-2014}. Exact diagonalization studies demonstrated that for typical 
observables $\sigma(\widehat{\mathcal O})\propto D^{-1/2}$~\cite{beugeling-2013}, 
with $D$ the dimension of the full Hilbert space. Oppositely, in integrable models 
a much slower decay is expected~\cite{deutsch-1991,rigol-2008}, which is associated with the 
presence of an extensive number of local conservation laws~\cite{caux-2011}. 
Yet, despite integrability, numerical studies of ETH are hampered by severe 
finite-size effects, and a precise finite-size scaling analysis of $\sigma(
\widehat{\mathcal O})$ is still lacking (see, however, Ref.~\onlinecite{ikeda-2013} 
for a finite-size scaling study of ETH in the Lieb-Liniger model).

%%%%%%%%%%%%%%%%%%%%%%%%%%%%%%%%%e
\begin{figure*}[t]
\includegraphics*[width=0.95\linewidth]{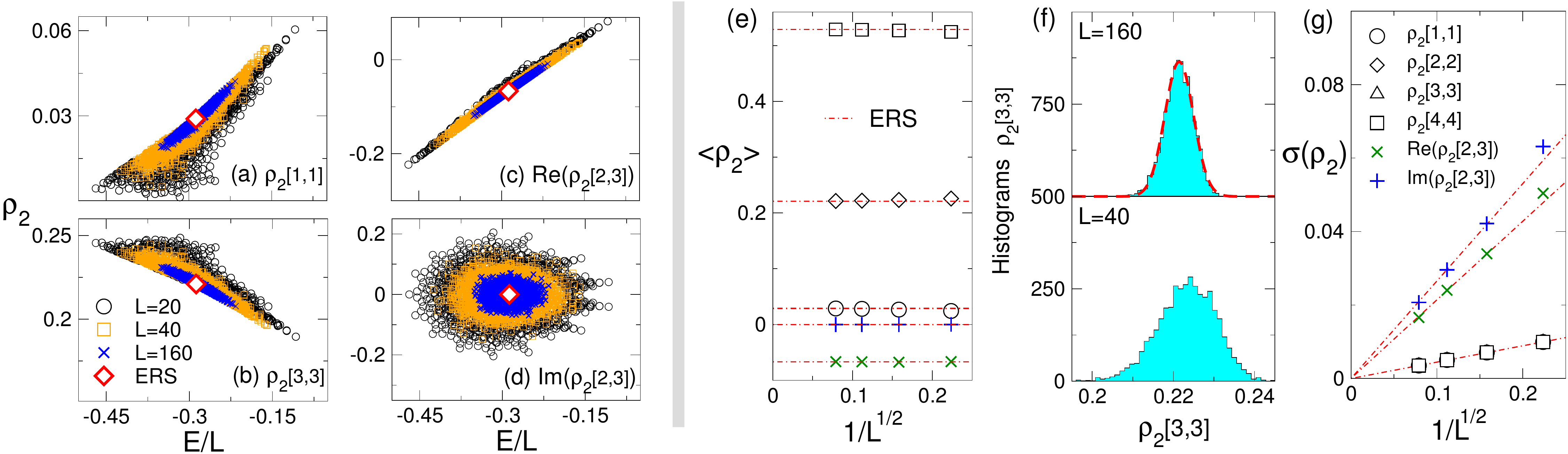}
\caption{
 Finite-size scaling of the ETH for the two-spins reduced
 matrix $\rho_{2}$ in the $XXX$ chain. (a)-(d) Matrix 
 elements $\rho_{2}[1,1],\rho_{2}[3,3]$, and $\rho_{2}[2,3]$ 
 plotted versus the eigenstates energy density $E/L$. 
 Panels (c) and (d) plot the real and imaginary part of $\rho_{2}
 [2,3]$, respectively. Circles, squares, and crosses denote chain 
 sizes $L=20,40,160$. The data correspond to $\sim 10^4$ eigenstates 
 obtained from real solutions of the Bethe equations at fixed particle 
 density $x\equiv M/L=1/4$, with $M$ the number of down spins 
 (particles). In all panels the rhombi denote the result obtained 
 from the ensemble representative state (ERS). All the matrix 
 elements are well described by the ERS in the limit $L\to\infty$. 
 (i) Eigenstate average $\langle\rho_{2}\rangle$ plotted versus $L^{-1/2}$. 
 Different symbols correspond to different matrix elements (cf. (iii)). 
 The dashed-dotted lines are the results obtained from the ERS. (ii) 
 Eigenstate-to-eigenstate fluctuations of $\rho_{2}$. Histograms of 
 $\rho_{2}[3,3]$ obtained from different eigenstates and $L=40,160$. 
 The histogram for $L=160$ is shifted vertically for visibility. 
 The dashed line is a gaussian fit. (iii) Fluctuations of $\rho_2$: 
 Standard deviation $\sigma(\rho_{2})$ plotted versus $L^{-1/2}$. 
 The dashed-dotted lines are linear fits. $\sigma(\rho_{2})$ is 
 vanishing in the limit $L\to\infty$. 
}
\label{fig1}
\end{figure*}
%%%%%%%%%%%%%%%%%%%%%%%%%%%%%%%%%%

%%% RESULTS 
\section{Summary of the results} 
In this work, using a numerical implementation (cf. Ref.~\onlinecite{alba-2009} 
for the details) of state-of-the-art Bethe ansatz results~\cite{kitanine-1999,
kitanine-2000}, we perform a finite-size scaling analysis of the ETH in the 
spin-$\frac{1}{2}$ isotropic Heisenberg ($XXX$) chain. Specifically, here we 
focus on the $\ell$-spin reduced density matrix $\rho_\ell$. Given the state of 
the chain $|\psi\rangle$ and a block $A$ of $\ell$ contiguous 
spins, the corresponding $\rho_\ell$ is obtained  by tracing over the degrees 
of freedom of the remaining spins (block $B$) in the full-system density 
matrix $\rho\equiv|\psi\rangle\langle\psi|$, i.e., $\rho_\ell\equiv\textrm{
Tr}_B\rho$~\cite{amico-2008}. From $\rho_\ell$ it is straightforward to 
obtain any multi-spin correlation function with support on $\ell$ contiguous 
sites of the chain. 

Our main result is that the weak ETH holds for all the matrix elements of 
$\rho_\ell$, meaning that for the vast majority of the eigenstates (i.e., 
typical eigenstates) one has $\rho_\ell\to\langle\rho_\ell\rangle$ (with 
$\langle\cdot\rangle$ denoting the microcanonical average), in the 
thermodynamic limit. This scenario is made rigorous by a large scale 
analysis of the eigenstate-to-eigenstate fluctuations of $\rho_\ell$. 
Precisely, we numerically demonstrate that fluctuations are normally 
distributed, and their standard deviation $\sigma(\rho_\ell)$ decays as 
$\sigma(\rho_\ell)\propto L^{-1/2}$, for large chains. 
Interestingly, the same behavior is observed in free models~\cite{biroli-2010}, 
and it is in sharp contrast with the exponential decay found in  
non-integrable systems~\cite{beugeling-2013}. Moreover, using standard 
thermodynamics Bethe ansatz (TBA) results~\cite{taka-book,yang-1969,takahashi-1971}, 
we provide an ensemble representative state (ERS), and the reduced density matrix 
$\rho_\ell^{ERS}$ thereof. This is a good approximation for the reduced density 
matrix obtained from typical mid-spectrum eigenstates of the finite chain. In 
particular, one has $\rho_\ell\to\rho_\ell^{ERS}$ in the thermodynamic limit. 
From $\rho_\ell$ we construct the conserved charges $I_n$ ($n\in{\mathbb 
N}$) of the $XXX$ chain~\cite{grabowski-1995}. Interestingly, we numerically 
observe that the EEV of the first non-trivial charge densities $I_n/L$ vanishes 
for typical eigenstates. One consequence of our results is that, 
despite the $XXX$ chain being integrable, the unitary dynamics ensuing from thermal-like 
initial states might lead to thermal behavior at long times, in agreeement with what 
has been found in Ref.~\onlinecite{santos-2012,santos-2012a}. Finally, we investigate the 
entanglement entropy~\cite{amico-2008,eisert-2009,calabrese-2009,cc-rev} 
$S_\ell\equiv\textrm{Tr}\rho_\ell\log\rho_\ell$ of the ERS. We provide robust 
numerical evidence for the volume law scaling $S^{ERS}_\ell\propto\ell$, which 
implies that typical mid-spectrum eigenstates exhibit extensive entanglement 
entropy. 

%%% MODEL AND METHOD
\section{The model and the method}
The spin-$\frac{1}{2}$ isotropic Heisenberg ($XXX$) chain is defined by the 
Hamiltonian 
\begin{align}
\label{xxx_ham}
{\mathcal H}\equiv J\sum\limits_{i=1}^L\left[\frac{1}{2}(S_i^+S^-_{i+1} 
+S_i^{-}S_{i+1}^+)+S_i^zS_{i+1}^z\right],  
\end{align}
where $S^{\pm}_i\equiv (\sigma_i^x\pm i\sigma_i^y)/2$ are the ladder 
operators acting on the site $i$ of the chain, $S_i^z\equiv\sigma_i^z/2$, 
and $\sigma^{x,y,z}_i$ the Pauli matrices. We set $J=1$ in~\eqref{xxx_ham} 
and use periodic boundary conditions identifying sites $L+1$ and $1$. 
Both the total spin $S_T^2\equiv(\sum_i \vec S_i)^2$ and the total 
magnetization $S_{T}^z\equiv\sum_iS_i^z=L/2-M$, with $M$ being the 
number of down spins (particles), are conserved quantities for 
${\mathcal H}$. Thus its eigenstates can be labeled by $S_T^z$ 
(equivalently by $M$, or the particle density $x\equiv M/L$). 
Here we mainly consider the situation with fixed density $x=1/4$.

In the algebraic Bethe ansatz approach~\cite{kor-book} the eigenstates 
of the $XXX$ chain in the sector with given $M$ are constructed from 
the so-called rapidities $\{\lambda_\alpha\}$ as $|\{\lambda_\alpha\}
\rangle\equiv\prod_{\beta=1}^MB(\lambda_\beta)|\Uparrow\rangle$. 
Here $|\Uparrow\rangle\equiv|\uparrow\uparrow\cdots\uparrow\rangle$ 
is the ferromagnetic (``vacuum'') state, and $B(\lambda)$ a $2^L
\times 2^L$ matrix~\cite{kor-book}. The rapidities are obtained 
by solving the Bethe equations 
\begin{equation}
\arctan(\lambda_\alpha)=\frac{\pi}{L}J_\alpha+
\frac{1}{L}\sum\limits_{\beta\ne \alpha}\arctan\Big(\frac{
\lambda_\alpha-\lambda_\beta}{2}\Big).   
\label{ba_eq}
\end{equation}
In principle, any choice of the so-called Bethe numbers $-L/2<J_\alpha
\le L/2$ ($J_\alpha\in \frac{1}{2}{\mathbb Z}$) identifies 
a set of solutions of~\eqref{ba_eq}, and an eigenstate of~\eqref{xxx_ham} 
thereof, with energy eigenvalue $E=\sum_{\alpha}2/(\lambda^2_\alpha
+1)$. Although, in general,  $\lambda_\alpha\in{\mathbb C}$, here we 
consider only real rapidities, i.e., $\lambda_\alpha\in{\mathbb 
R},\,\forall\alpha$. Moreover, we focus on the eigenstates with 
maximum magnetization $S_T^z=S_T$, which implies $\lambda_\alpha<\infty
\,,\forall\alpha$~\cite{taka-book}. The corresponding Bethe numbers are 
given as $-J_{\infty}\le J_\alpha\le J_{\infty}$, with $J_{\infty}=(L-1-
M)/2$~\cite{taka-book}. Finally, for each $L$ we consider $\sim 10^4$ 
eigenstates of~\eqref{xxx_ham}, which  are obtained by sampling uniformly 
the Bethe numbers $\{J_\alpha\}$. We should mention that here we choose 
$\Delta e\to\infty$, meaning  that we are considering the infinite 
temperature canonical ensemble, instead of a microcanonical one. However, 
we anticipate that this does not affect our results. 

Formally, for any eigenstate $|\{\lambda_\alpha\}\rangle$    
one can write the corresponding reduced density matrix $\rho_\ell$ as 
($\langle\{\lambda_\alpha\}|\{\lambda_\alpha\}\rangle=1$)  
\begin{equation}
\label{rdm_el}
\rho_{\ell}=\langle\{\lambda_\alpha\}|\prod\limits_{j=1}^\ell 
\sigma_j^{\varepsilon_j}|\{\lambda_\alpha\}\rangle,  
\end{equation}
where $\varepsilon_j=x,y,z,0$, and $\sigma^0\equiv{\mathbb I}$ is 
the $2\times 2$ identity matrix. 
Exact formulas for the multi-spin correlation functions appearing 
in~\eqref{rdm_el} have been obtained recently within the algebraic 
Bethe ansatz approach~\cite{kitanine-1999,kitanine-2000}. Their numerical 
implementation is a challenging task (cf. Ref.~\onlinecite{alba-2009} 
for details), and the computational cost for calculating a generic matrix 
element of $\rho_\ell$ is roughly $\propto L^\ell$. Nonetheless, this 
method allows one to effectively calculate $\rho_{\ell}$ for $\ell\lesssim 
6$ and chains with $L\sim 100$. We mention that this has been used in 
Ref.~\onlinecite{alba-2009} to study entanglement properties of the 
excited states of the $XXZ$ chain.

%%%%%%%%%%%%%%% REDUCED-DENSITY-MATRIX ELEMENTS %%%%%%%%%%%%%%%%%

\section{The fluctuations of the reduced density matrix}

%%%%%%%%%%%%%%%%%%%%%%%%%%%%%%%%%e
\begin{figure}[t]
\includegraphics[width=0.95\columnwidth]{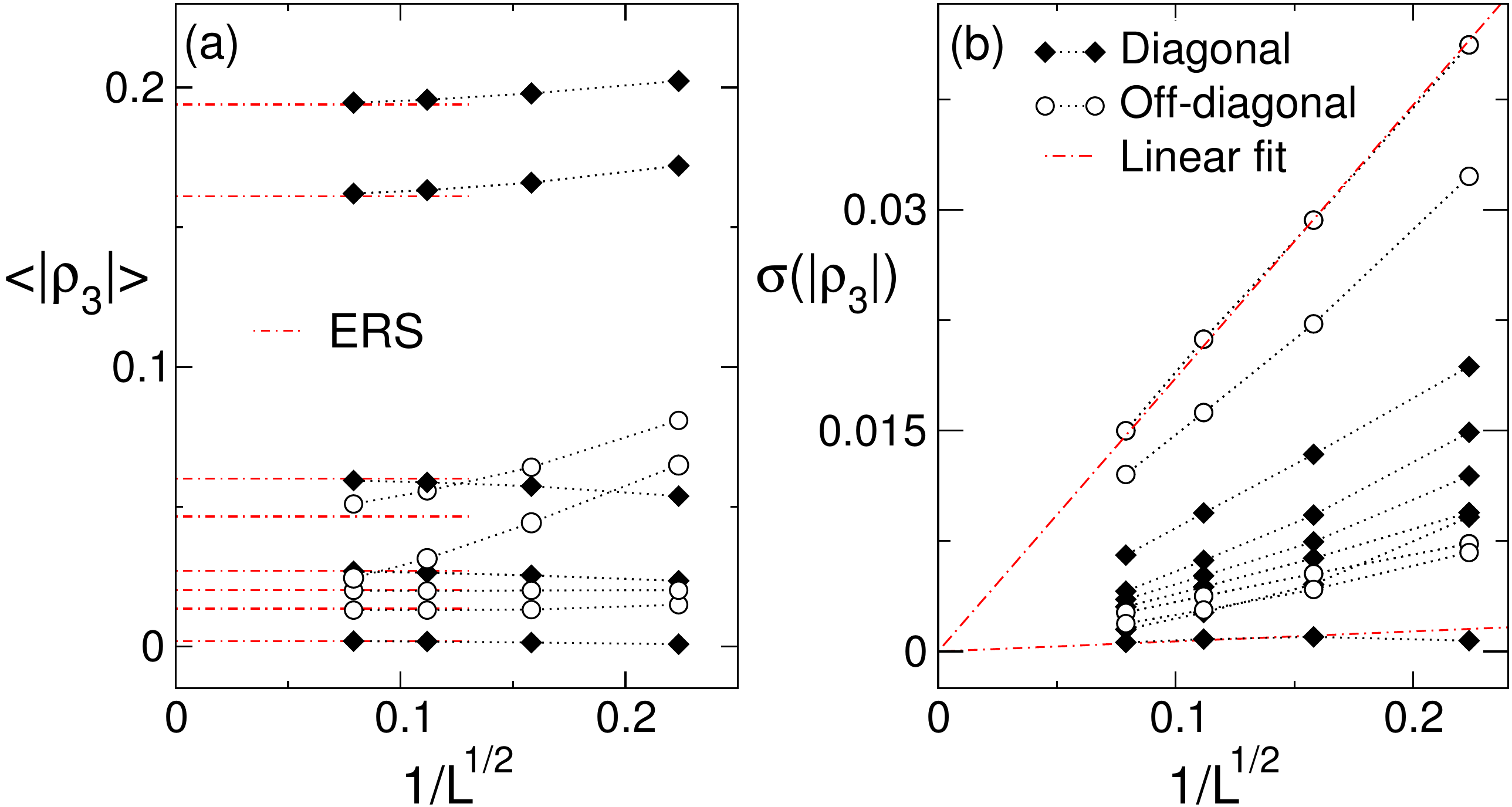}
\caption{
 Finite-size scaling of the eigenstate-to-eigenstate fluctuations 
 of the three-spins reduced density matrix $\rho_{3}$ in the $XXX$ 
 chain. (a) Eigenstate average $\langle |\rho_{3}|\rangle$ of $\rho_{3}$ 
 plotted against $L^{-1/2}$ for chains with $L=20,40,80,160$. The 
 average is over $\sim 10^4$ eigenstates in the sector with fixed 
 particle density $x\equiv M/L=1/4$, $M$ being the number of down 
 spins (particles). Full and empty symbols denote the diagonal 
 and off-diagonal matrix elements of $\rho_3$, respectively. The 
 dash-dotted lines denote $\rho_3^{ERS}$. (b) Fluctuations of 
 $|\rho_{3}|$: Standard deviation $\sigma(|\rho_{3}|)$ plotted 
 versus $L^{-1/2}$. The dash-dotted lines are linear fits. 
 The data suggest that $\sigma(|\rho_{3}|)\to 0$ in the 
 thermodynamic limit.   
}
\label{fig3}
\end{figure}
%%%%%%%%%%%%%%%%%%%%%%%%%%%%%%%%%%

Our main results are illustrated in Figure~\ref{fig1} focusing on 
$\rho_2$. Figure~\ref{fig1} (a) and (b) plot its diagonal elements 
$\rho_{2}[1,1]$ and $\rho_{2}[3,3]$ versus the eigenstate energy 
density $E/L$. We verified that $\rho_2[2,2]$ and $\rho_2[4,4]$ 
exhibit  the same qualitative behavior. The only non-zero off-diagonal 
element $\rho_{2}[2,3]=(\rho_{2}[2,3])^*$ is in general complex, 
and its real and imaginary parts are plotted in (c) and (d), 
respectively. In all panels the different symbols correspond to 
the chain lengths $L=20,40,160$ at fixed particle density $x=1/4$. 

\begin{figure}[t]
\includegraphics[width=0.95\columnwidth]{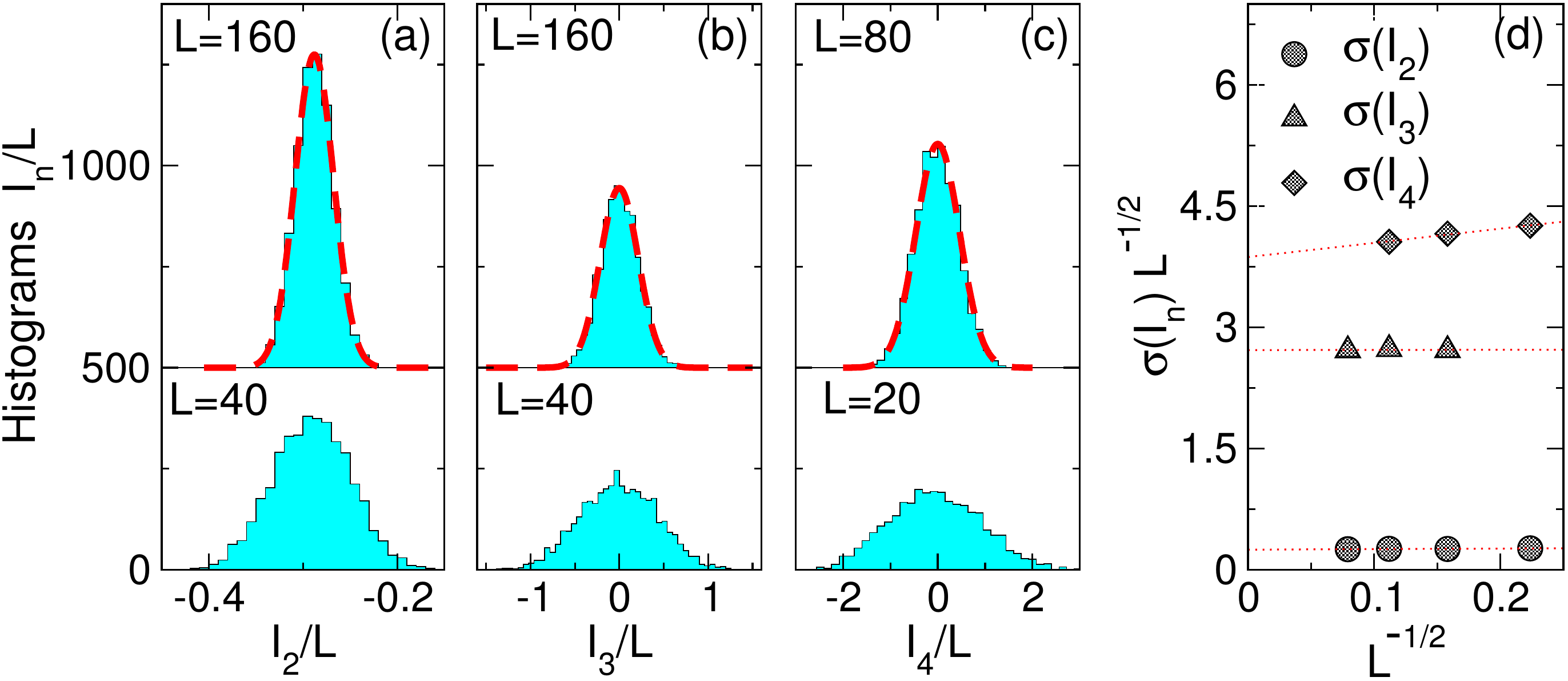}
\caption{
 Eigenstates expectation values (EEV) of the conserved charges 
 $I_n$ ($n=2,3,4$) of the $XXX$ chain. (a)-(c) Histograms of 
 $I_n/L$ for several chain lengths $L=20,40,80,160$. 
 Here $I_2$ and $I_3$ are the energy $E$ and the energy current 
 $J_E$, respectively. For each $L$ the data are obtained 
 from $\sim 10^4$ eigenstates with fixed particle density $x\equiv 
 M/L=1/4$, $M$ being the number of down spins. The histograms 
 for the larger chains are shifted vertically for visibility. The 
 dashed lines are gaussian fits. In the limit $L\to\infty$ one has 
 $I_2/L\approx -0.3$, whereas $I_3/L\approx I_4/L\approx 0$. (d) 
 Rescaled fluctuations $\sigma(I_n)L^{-1/2}$ plotted versus 
 $L^{-1/2}$. The dotted lines are linear fits. Notice the 
 finite extrapolations in the thermodynamic limit.  
}
\label{figx}
\end{figure}

In the limit $L\to\infty$ the vast majority of the eigenstates exhibit  
the typical energy density $e_{typ}\equiv E_{typ}/L\approx -0.2876\cdots$ 
(cf. Figure~\ref{figx} for a precise analysis). This suggests, as expected, 
that for large $L$ the infinite temperature canonical ensemble can be replaced 
with a microcanonical one centered around $e_{typ}$. Correspondingly, 
Figure~\ref{fig1} (a)-(d) demonstrate that the eigenstate-to-eigenstate 
fluctuations of $\rho_{2}$ decrease upon increasing $L$, which implies that 
in the thermodynamic limit typical eigenstates yield the same $\rho_{2}$, 
and one has $\rho_2\to\langle\rho_2\rangle$. This allows us to conclude 
that the weak ETH holds for $\rho_2$. Finally, it is interesting to observe 
that, while in general $\rho_{2}[2,3]\in{\mathbb C}$, one has $\textrm{Im}(
\rho_{2}[2,3])\approx 0$ at $L\to\infty$. 

All these findings can be better characterized within the TBA approach 
by constructing an ensemble representative state (ERS) for the typical 
eigenstates. In the TBA, instead of the eigenstates, one considers the 
rapidity (particle) densities $\varrho_p(\lambda)$, whereas sums over 
eigenstates are replaced by a functional integral over $\varrho_p(
\lambda)$ as~\cite{taka-book} 
\begin{equation}
|\{\lambda_\alpha\}\rangle\to\varrho_p(\lambda),\quad 
\sum\limits_{|\{\lambda_\alpha\}\rangle}\to\int{\mathcal D}[\varrho_p]
e^{S_{YY}[\varrho_p]}. 
\label{tba}
\end{equation}
Here $\varrho_h$ denotes the ``hole'' (i.e., missing rapidities) density, and 
$S_{YY}[\varrho_p]\equiv L\int d\lambda[(\varrho_p+\varrho_h)\log(\varrho_p+
\varrho_h)-\varrho_p\log\varrho_p-\varrho_h\log\varrho_h]$ is the so-called 
Yang-Yang entropy. In the thermodynamic limit the integral in~\eqref{tba} is 
dominated by the saddle point $\bar\varrho_p$ of $S_{YY}[\varrho_p]$. At fixed 
density $x$, by imposing $\delta S_{YY}/\delta\varrho_p|_{\bar\varrho_p}=0$, one 
obtains that the ratio $\bar\varrho_h/\bar\varrho_p$ does not depend on $\lambda$, 
and it is given as $\bar\varrho_h/\bar\varrho_p=(1-2x)/x$. Moreover, it is 
straightforward to derive from the Bethe equations that $2\pi\bar\varrho_p(
\lambda)=\int_{-\infty}^{+\infty}d\omega e^{-|\omega|+i\omega\lambda}/[(1-x)/x
+e^{-2|\omega|}]$. The corresponding energy density, i.e., $e_{typ}$,  
is readily obtained as $e_{typ}=2\int d\lambda\bar\rho_p/(1+\lambda^2)$~\cite{taka-book}. 
Notice that the peaking of the eigenstates energy density around $e_{typ}$ (cf. 
Figure~\ref{fig1} (a)-(d) and Figure~\ref{figx} (a)) in the thermodynamic limit is 
a consequence of $S_{YY}$ being extensive (cf.~\eqref{tba}). 

For a finite chain, the condition $\bar\varrho_h/\bar\varrho_p=(1-2x)/x$ implies 
that the saddle point is well approximated  by the eigenstate (ERS) corresponding  
to the Bethe numbers $J_\alpha=-(1-x)/2L+1/2+(1/x-1)\alpha$, with $\alpha=0,1,\dots, M-1$. 
Interestingly, this choice maximizes the ``sparseness'' of the Bethe numbers. 
The matrix elements of the resulting $\rho_{2}^{ERS}$ (for a chain with $L=80$) 
are reported in Figure~\ref{fig1} (a)-(d) as rhombi. For large chains one has 
$\rho_{2}\to\rho_{2}^{ERS}$ for typical eigenstates, which implies $\langle
\rho_{2}\rangle\to\rho_2^{ERS}$, as shown in Figure~\ref{fig1} (i). 

We now turn to discuss the fluctuations of $\rho_2$. These are illustrated in 
Figure~\ref{fig1} (ii), plotting the histograms of $\rho_2[3,3]$ for $L=40$ 
and $L=160$  (the latter is shifted vertically for visibility). Similar results 
are obtained for other matrix elements (not shown). The dashed line is a gaussian 
fit, suggesting that the fluctuations are normally distributed, as for non-integrable 
models~\cite{beugeling-2013}. Their amplitude vanishes upon increasing $L$, as 
expected (cf. Figure~\ref{fig1} (a)-(d)). This is numerically demonstrated for all 
matrix elements of $\rho_2$ in Figure~\ref{fig1} (iii), by plotting $\sigma(\rho_2)$ 
versus $L^{-1/2}$. Clearly, one has $\sigma(\rho_{2})\propto L^{-1/2}$ (dash-dotted 
lines are linear fits). 

One should remark that it is straightforward to derive that $\rho_{2}[1,1]=(1+
4G^{zz})/4$, with $G^{zz}\equiv\langle S_i^zS_{i+1}^z\rangle=-\langle {\mathcal 
H}/L\rangle/3$, where in the last step the $SU(2)$ invariance of ${\mathcal H}$ 
was used. This suggests that the weak ETH scenario for $\rho_2$ could be a trivial 
consequence of the behavior of the eigenstate energy density (cf. Figure~\ref{figx} 
(a)). In order to provide a more stringent check of our result, in Figure~\ref{fig3} 
we discuss $\rho_{3}$. Figure~\ref{fig3} (a) plots all non-zero elements of $\langle|
\rho_{3}|\rangle$ versus $L^{-1/2}$. Full and empty symbols denote diagonal and 
off-diagonal elements, respectively. The matrix elements of $\rho_{3}^{ERS}$ are shown 
as dash-dotted lines. Although finite-size effects are larger compared with 
$\rho_{2}$ (cf. Figure~\ref{fig1} (i)), one has $\langle\rho_{3}\rangle\to\rho_{3}^{ERS}$ 
at $L\to\infty$. The eigenstate-to-eigenstate fluctuations of $\rho_{3}$ are investigated 
in Figure~\ref{fig3} (b) plotting $\sigma(|\rho_{3}|)$ versus $L^{-1/2}$. Clearly, 
$\sigma(|\rho_{3}|)\propto L^{-1/2}$ is vanishing in the thermodynamic limit, as for 
$\rho_{2}$ (cf. Figure~\ref{fig1} (iii)). Finally, we should mention that similar 
qualitative behavior is observed for $\rho_4$.

%%%%%%%%%%%%%%%%%%%%%%%%%%%%%%%%%
%%%%%%%%%%%%%%%%%%%%%%%%%%%%%%%%%%

%%% CONSERVED CHARGES 
\section{The conserved charges} 
It is interesting to consider 
the EEVs of the {\it local} conserved charges $I_n$ of the $XXX$ 
chain~\cite{grabowski-1995}. For each $n$, these are obtained as    
particular linear combinations of the matrix elements of $\rho_n$. 
Figure~\ref{figx} plots the histograms of $I_n/L$ for $n=2,3,4$, 
(panels (a)-(c)) and $L=20,40,80,160$. Some of the histograms are 
shifted vertically for visibility. Here $I_2\equiv{\mathcal H}$, 
while $I_3$ is the energy current $I_3=J_E\equiv\sum_{\alpha\beta
\gamma}\epsilon_{\alpha\beta\gamma}\sigma^\beta_i\sigma_{i+1}^\gamma
\sigma_{i+2}^\alpha$, with $\alpha,\beta,\gamma=x,y,z$ and $\epsilon_{
\alpha\beta\gamma}$ the Levi-Civita symbol. The EEVs of $I_n/L$ are 
normally distributed around the typical values $\bar I_n/L\equiv\langle 
I_n/L\rangle$, as confirmed by the gaussian fits (dashed lines in the 
Figure). Interestingly, while $\bar I_2/L\approx -0.2876\cdots$ (panel 
(a), see also Figure~\ref{fig1}), one has $\bar I_3/L\approx
\bar I_4/L\approx 0$ (panels (b) and (c)). The fluctuations of $I_n/L$ 
decay as $L^{-1/2}$  in the thermodynamic limit, reflecting the behavior 
of $\rho_\ell$ (cf. Figure~\ref{fig1} and~\ref{fig3}). This is verified  
in Figure~\ref{figx} (d) plotting $\sigma(I_n)L^{-1/2}$ versus $L^{-1/2}$. 
Clearly, $\sigma(I_n)L^{-1/2}\to const$ at $L\to\infty$ (the dotted lines 
are linear fits). Physically, the behavior of $I_2$  reflects the specific 
heat per site, $C_v=\beta^2/L(\langle I_2^2\rangle-\langle I_2\rangle^2)$, 
being finite in the thermodynamic limit. Likewise, the result for $I_3$ 
is related to the finite Drude weight for the energy current~\cite{zotos-1996,
castella-1996,zotos-1997}.

%%%%%%%%%%%%%%%%%%%%%%%%%%%%%%%%%e
\begin{figure}[t]
\includegraphics[width=0.95\columnwidth]{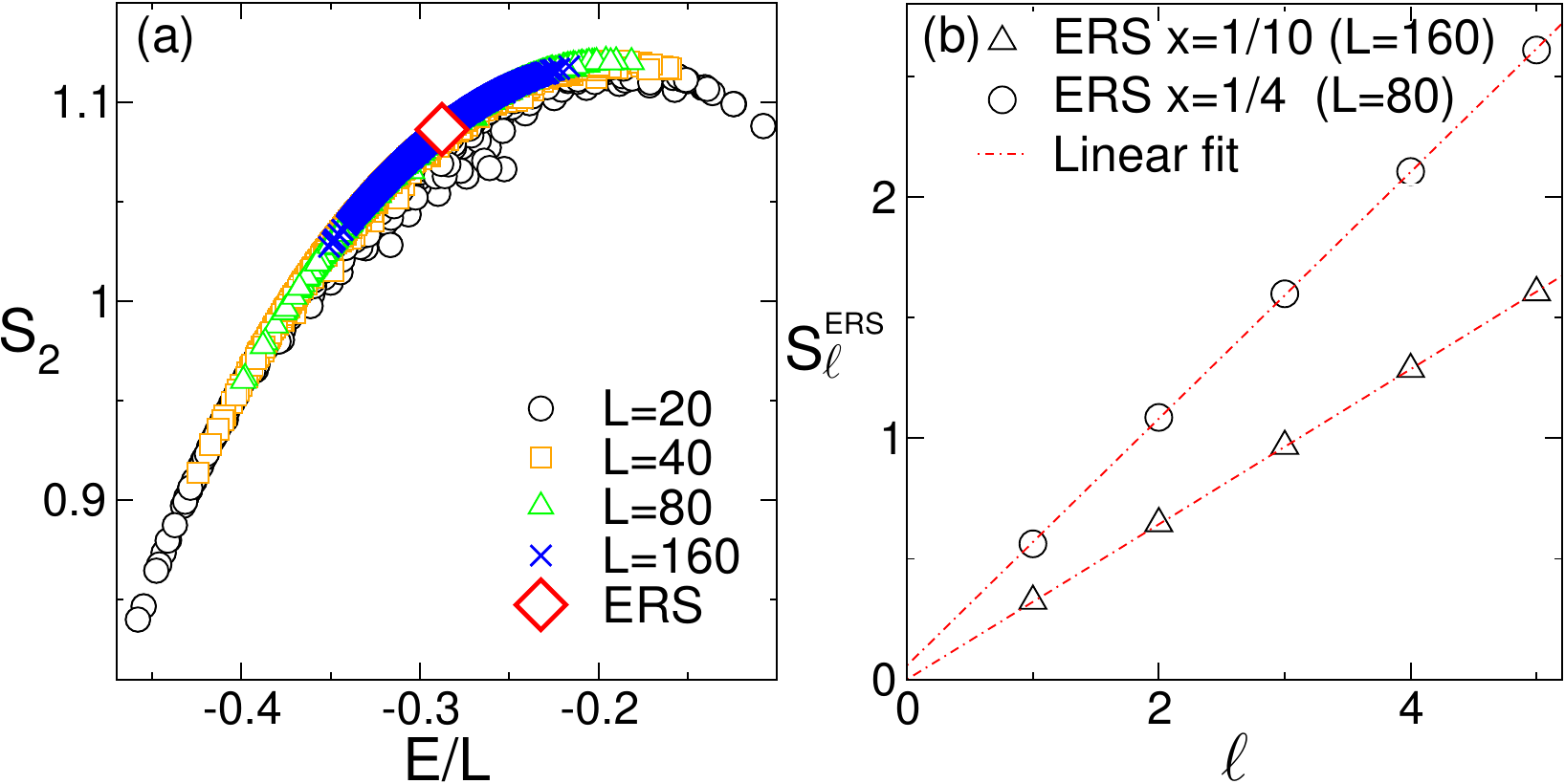}
\caption{
 Entanglement entropy of the eigenstates of the spin-$\frac{1}{2}$ 
 $XXX$ chain. (a) Two-spin von Neumann entropy $S_2$ plotted 
 versus the eigenstates energy density $E/L$. Different 
 symbols correspond to chain lengths $L=20,40,80,160$. The 
 rhombus denotes $S^{ERS}_2$. (b) $S^{ERS}_\ell$ as a function 
 of  $\ell=1,\dots, 5$, and for $L=80,160$. Circles and triangles 
 correspond to particle densities $x\equiv M/L=1/4,1/10$, 
 respectively. Here $M$ is the number of down spins (particles). 
 The dashed-dotted lines are linear fits. For both $x=1/4,1/10$ 
 the volume law $S_\ell^{ERS}\propto\ell$ is clearly visible. 
}
\label{fig4}
\end{figure}
%%%%%%%%%%%%%%%%%%%%%%%%%%%%%%%%%%

%%% ENTANGLEMENT PROPERTIES 

\section{Entanglement properties} 
We now turn to discuss how the scenario outlined so far is reflected 
in the entanglement entropy $S_\ell$ of the eigenstates of the $XXX$ 
spin chain. Notice that entanglement properties of excited states 
of many-body Hamiltonians have attracted increasing attention 
recently~\cite{alba-2009,alcaraz-2011,pizorn-2012,berganza-2012,
wong-2013,storms-2013,berkovits-2013,essler-2013,nozaki-2014,ramirez-2014,
ares-2014,huang-2014,palmai-2014,molter-2014,lai-2014}. Figure~\ref{fig4} (a) 
plots the two-spin entropy $S_2$ versus $E/L$ for different eigenstates of 
the $XXX$ chain. The eigenstate-to-eigenstate fluctuations of $S_2$ decrease 
upon increasing $L$, reflecting the behavior of $\rho_{2}$ (cf. Figure~\ref{fig1}). 
In particular, in the thermodynamic limit one has $S_2\to S_2^{ERS}$. The scaling 
behavior of $S_\ell^{ERS}$, as a function of $\ell$ is investigated in 
Figure~\ref{fig4} (b) plotting $S^{ERS}_\ell$ versus $1\le\ell\le 5$ for chains 
of length $L=80,160$. We restrict 
ourselves to density $x=1/4$ (high density) and $x=1/10$ (low density). In both 
cases Figure~\ref{fig4} (b) provides a robust evidence of the volume law 
$S_\ell^{ERS}\sim\ell$ (the dash-dotted lines are linear fits). This allows 
us to conclude that typical mid-spectrum eigenstates of the $XXX$ chain exhibit 
extensive entanglement (see also Ref.~\onlinecite{sato-2011}), whereas other 
behaviors (for instance a logarithmic one as $\propto\log(\ell)$) are associated 
with rare eigenstates. Notice that this should be reflected in the extensive 
behavior of the entanglement entropy after a quantum quench~\cite{fagotti-2008,
gurarie-2013,fagotti-2013a,collura-2014,kormos-2014}.

\section{Conclusions} 
We performed a careful finite-size scaling analysis of the eigenstate 
thermalization hypothesis (ETH) in the spin-$\frac{1}{2}$ isotropic Heisenberg 
($XXX$) chain. Precisely, we focused on the $\ell$-spin reduced density matrix 
$\rho_\ell$. Using a numerical implementation~\cite{alba-2009} of state-of-the-art 
algebraic Bethe ansatz results~\cite{kitanine-1999,kitanine-2000}, we provided 
numerical evidence that ETH holds for typical eigenstates ({\it weak} ETH 
scenario). Specifically, we numerically demonstrated that the eigenstate-to-eigenstate 
fluctuations of $\rho_\ell$ decay as $L^{-1/2}$ in the thermodynamic limit. This is 
in contrast with the exponential decay that is observed in generic non-integrable 
models~\cite{beugeling-2013}, for which the {\it strong} ETH holds. Although here we 
considered one specific model, and we restricted ourselves to $\rho_\ell$, it is 
natural to expect  that the same result holds in other integrable spin systems and 
for generic {\it local} observables. Finally, we numerically verified that typical 
mid-spectrum eigenstates of the $XXX$ chain exhibit extensive entanglement entropy.  

Our results open several new possible research avenues. For instance, while here 
we restricted ourselves to eigenstates obtained from real solutions of the Bethe 
equations, it would be interesting to generalize our results including complex 
solutions (strings)~\cite{taka-book}. Notice that for $\ell=2$ this should be 
straightforward using the techniques developed in Ref.~\onlinecite{caux-2005,
caux-2005a,caux-2009}. Furthermore, it would be useful to characterize the scaling 
behavior of the entanglement entropy in eigenstates away from the middle of the 
energy spectrum. An intriguing direction would be to investigate the relationship 
between entanglement and the spectrum of the conserved charges (other than the 
Hamiltonian). In particular, it would be enlightening to study whether eigenstates 
at the edges of the spectrum give non-extensive entanglement entropy. Finally, it 
would interesting to extend our analysis to off-diagonal matrix elements of local 
observables, which are important to understand the approach to relaxation to a 
steady state after quantum quenches~\cite{khatami-2013,beugeling-2014}. 

%%% ACKNOWLEDGMENTS 
\paragraph*{Acknowledgments.---}
I would like to thank P.~Calabrese, F.~H.~L.~Essler, M.~Fagotti, 
F.~Heidrich-Meisner, A.~M.~L\"auchli, and A.~Lazarides for 
enlightening discussions and comments. In particular, I would like to 
thank F.~H.~L.~Essler for many useful insights during the completion 
of this work and P.~Calabrese and M.~Fagotti for useful comments about 
the manuscript. I acknowledge financial support by the ERC under Starting 
Grant 279391 EDEQS.

\end{document}